\documentclass[preprint]{aastex}
\usepackage{color}
\usepackage{psfig}
\newcommand{\lta}{$\; \buildrel < \over \sim \;$}
\newcommand{\simlt}{\lower.5ex\hbox{\lta}}
\newcommand{\gta}{$\; \buildrel > \over \sim \;$}
\newcommand{\simgt}{\lower.5ex\hbox{\gta}}

\newcommand{\lsun}{{\rm\,L_\odot}}

\newcommand{\ffffff}[1]{\mbox{$#1$}}
\newcommand{\scnd}{\mbox{\ffffff{''}\hskip-0.3em.}}
\newcommand{\scmd}{\mbox{\ffffff{''}}}

\newcommand{\apm}{APM~08279+5255}

\slugcomment{Accepted by the P.A.S.P.}
\shortauthors{G. F. Lewis et. al.}
\shorttitle{Photometric Monitoring of \apm}
\begin{document}

\title{Photometric Monitoring of the Gravitationally Lensed 
Ultraluminous \\ BAL Quasar \apm}

\author{
Geraint F.  Lewis\altaffilmark{1,2,3,4}, 
Russell M. Robb\altaffilmark{3,5} \&
Rodrigo A. Ibata\altaffilmark{6,7}}

\altaffiltext{1}{
Astronomy Dept., University of Washington, Box 351580, 
Seattle WA 98195-1580, U.S.A.}

\altaffiltext{2}{
gfl@astro.washington.edu}

\altaffiltext{3}{
Dept. of Physics and Astronomy, University of Victoria, 
PO Box 3055,  Victoria, B.C.,  V8W 3P6, Canada}

\altaffiltext{4}{
gfl@uvastro.phys.uvic.ca}

\altaffiltext{5}{
robb@uvastro.phys.uvic.ca}

\altaffiltext{6}{
European Southern Observatory, Garching bei M\"unchen, Germany}

\altaffiltext{7}{
ribata@eso.org}

\begin{abstract}
We report on  one year of photometric monitoring  of the ultraluminous
BAL   quasar  \apm.    The   temporal  sampling   reveals  that   this
gravitationally  lensed  system has  brightened  by  $\sim$0.2 mag  in
100~days.   Two  potential   causes  present  themselves;  either  the
variability  is  intrinsic to  the  quasar, or  it  is  the result  of
microlensing by stars in a  foreground system.  The data is consistent
with both hypotheses and  further monitoring is required before either
case  can be  conclusively confirmed.   We demonstrate,  however, that
gravitational microlensing can not  play a dominant role in explaining
the phenomenal  properties exhibited  by \apm.  The  identification of
intrinsic variability,  coupled with the  simple gravitational lensing
configuration, would  suggest that \apm\ is a  potential `golden lens'
from which the cosmological parameters can be derived and is worthy of
a monitoring program at high spatial resolution.
\end{abstract}


\section{Introduction}\label{introduction}
During a survey of carbon stars in the Galactic halo~\citep{totten98},
spectroscopic observations of a stellar  candidate revealed it to be a
broad   absorption  line  quasar   at  a   redshift  of   ${\rm  z\sim
3.9}$~\citep{irwin98}.  With  ${\rm m_R=15.2}$ this  system apparently
possesses  a  luminosity exceeding  ${\rm  10^{15}\lsun}$, placing  it
amongst  the  most  luminous  objects currently  known.   The  optical
emission  is  coincident  with  a  bright IRAS  source  and  follow-up
observations with Submillimetre Common-User Bolometer Array (SCUBA) on
the James Clerk Maxwell Telescope reveal that a substantial proportion
of this emission arises  in the sub-mm/IR regime~\citep{lewis98}. This
is characteristic of emission from  warm dust and indicates that \apm\
represents  one  of the  most  extreme  examples  of an  ultraluminous
infrared galaxy~\citep{sanders96}.
 
Analysis  of ground-based  images,  obtained with  the  1.0 m  Jacobus
Kapteyn Telescope on La Palma, revealed that the point spread function
of \apm\ was not stellar, but in fact was better represented by a pair
of  point-like  sources  separated  by  $\sim0\scnd4$~\citep{irwin98}.
Such  a configuration  is indicative  of the  action  of gravitational
lensing.  This conclusion was confirmed using adaptive optical imaging
at  the Canada-France-Hawaii  Telescope~\citep{ledoux98},  as well  as
images from  the 10m Keck telescope~\citep{egami99}  and NICMOS images
from the  Hubble Space Telescope~\citep{ibata99},  that clearly reveal
the  composite  nature of  this  system.   Several gravitational  lens
models  have been  derived  from these  studies,  suggesting that  the
magnification of  the quasar source is  $\sim10-100$.  Even accounting
for  this substantial  magnification, \apm\  remains amongst  the most
intrinsically   luminous  systems   currently  known.    Its  apparent
brightness, however, has  made it an ideal target  for several studies
of  high  redshift  quasars~(Downes  et al.~1999;  Hines,  Schmidt  \&
Smith~1999; Ellison et al.~1999; Ellison et al.~1999a)

In  this paper,  we too  take advantage  of the  gravitational lensing
magnification  and present  the  results of  one  year of  photometric
monitoring of  this phenomenal object.   In Section~\ref{observations}
we outline  the observational programme, data  reduction technique and
results  of  the  monitoring,  while in  Section~\ref{variability}  we
discuss     the     source     of    the     observed     variability.
Section~\ref{conclusions} presents  the conclusions of  this study and
discusses their relevance.

\section{Observational Programme}\label{observations}

\subsection{Observations and Data Reduction}
The 0.5-m telescope,  Cousins R filter and `STAR I'  CCD camera of the
Climenhaga  Observatory  of  the   University  of  Victoria  (Robb  \&
Honkanen, 1992)  were used  for the photometric  observations.  \apm's
apparent brightness makes it an  ideal extragalactic target for such a
telescope.   At a  redshift of  3.911, the  R-band  filter encompasses
${\rm Ly_\alpha~\lambda1216}$ and \ion{N}{5}~$\lambda1240$, as well as
weak  emission  of  Si+O~$\lambda1400$, with  \ion{C}{4}~$\lambda1909$
occurring in the red tail of  the filter [see Figure~1 of Irwin et al.
(1998)].  The prominent emission lines are strongly absorbed from both
self-absorption  and strong  broad absorption  line features,  and the
dominant flux in the band is continuum emission from the quasar.

Using standard  IRAF\footnote{IRAF is distributed  by National Optical
Astronomy  Observatories,  which is  operated  by  the Association  of
Universities for  Research in Astronomy,  Inc., under contract  to the
National   Science  Foundation}   routines,  the   frames   were  bias
subtracted,  flat-fielded,  and   the  magnitudes  were  derived  from
$6\scmd$ apertures using the ``Centroid'' centering option of the PHOT
package  [see~\citet{robb97}   for  more  details   of  the  reduction
procedure].  Due  to the small  field of view, extinction  effects are
negligible  and no  corrections have  been  made for  them.  Also,  no
corrections  were applied  to  transform the  R-band  magnitudes to  a
standard system.

The  field around  \apm\ is  shown  in Figure  1. In  addition to  the
quasar, several prominent stars are apparent.  Differential magnitudes
were calculated  by comparison with a bright  `standard' star labelled
{\bf  S2}.  The  automatic  observing procedure  exposed  a number  of
images of the  field, between 2 and 100, over  a night, dependent upon
weather   conditions.   Typically  $\sim50$   frames,  each   with  an
integration time  of 222 sec, were thereby  obtained.  The photometric
data in each of the 23 nights for which data was procured was examined
for brightness  variations that might have occurred  during the night,
but  no significant  variations  could  be attributed  to  any of  the
targets.  The  fields acquired  in a single  night of  observing were,
therefore, combined.

\subsection{Results}
Differential magnitudes  with respect to {\bf S2}  were calculated for
\apm\ and the other stars labelled in the Figure~1.  These results are
tabulated  in Table~1 and  are presented  graphically in  Figure~2. To
improve clarity, the light curves in this figure have been offset from
their original  values.  An  examination of the  light curves  for the
brighter stars in the field,  specifically {\bf S1}, {\bf S2} and {\bf
S4},  reveals that  they  have remained  constant  over the  observing
period.  This  also appears to be  the case for {\bf  S3}, although it
does possess several systematic variations that suggest variability.

The mean and standard deviation of these nightly means was found to be
$3.928\pm0.023$  for  {\bf  S1}-{\bf  S2}.   This  standard  deviation
assures us that both the stars  {\bf S1} and {\bf S2} are not variable
at this level of precision.  Comparing \apm\ to {\bf S2}, however, the
overall  differential  magnitude  was  found  to  be  $3.997\pm0.068$,
suggesting that  significant variability took place  during the course
of our monitoring campaign.

This is very apparent on examination of the light curves with the most
dramatic systematic change in  any light curve occurring in APM08-{\bf
S2}. While initially remaining constant  at the start of the observing
period,  at  Day~1230 it  begins  to  brighten,  with an  increase  of
$\sim0.2$ mag by the final observation.

\section{Variability}\label{variability}
What  is  the  source  of  the  observed  variability  in  \apm?   Two
possibilities  immediately  present  themselves, each  with  different
consequences  for our  understanding of  this  gravitationally lensed,
ultraluminous system.

\subsection{Intrinsic Variability}
While  some quasars show  spectacular multi-wavelength  variability on
short timescales  [e.g. Optically-Violent Variables~\citep{webb90} and
Intra-Day Variables~\citep{ked97}],  all quasars  are seen to  vary by
some degree  and several groups have employed  a variability selection
criterion      to      successfully      identify      samples      of
quasars~\citep{hawkins93}.   Thought to  arise  in variable  accretion
rates  and disk  instabilities in  the central  regions of  the quasar
nucleus, several studies of  the characteristics of quasar variability
have           been           undertaken           in           recent
years~\citep{di96,cristiani96,scholz97,Cristiani97,giveon99}  with the
goal   of  unraveling   the   structure  at   the   core  of   quasars
[c.f.~\citet{aretxaga97}].   \citet{hook94}  considered  a  sample  of
three hundred quasars  over an extended baseline of  several years and
found that, while  the amplitude of quasar variability  appears not to
be a function of redshift, there  is an inverse trend of the amplitude
of  the  variability  with  the  absolute  magnitude  of  the  quasar.
Considering  the  scatter in  their  relation,  however, the  observed
variability in  \apm\ is quite  consistent with it being  intrinsic to
the quasar.

\subsection{Gravitational  Microlensing}   
Given  the small  image separation,  the  light from  \apm\ must  pass
through  the inner  regions of  the lensing  galaxy where  the optical
depth to microlensing can be substantial.  Unlike the simple single or
binary star  microlensing seen in  the Galactic halo~\citep{alcock93},
microlensing in this regime results  in complex light curves which are
characterized   by   regions   of  high   magnification   variability,
interspaced         with          quiescent         periods         of
demagnification~\citep{kayser86}.   Could the observed  variability be
due to such microlensing?

The quintessential  example of a  microlensed quasar is  the quadruply
imaged Q2237+0305 which, with over  a decade of monitoring, has reveal
microlensing   variations  of   up  to   0.3  mag   in  a   matter  of
weeks~\citep{corrigan91,ostensen96}.   Comparing the  light  curves of
the  two systems  and noting  that, given  the  relative gravitational
lensing  geometry, the  time scale  of microlensing  events  should be
$\sim4\times$ longer than  those seen in Q2237+0305 (a  value which is
reasonably insensitive to the assumed cosmology), the observed rise is
also consistent with the microlensing hypothesis.

\section{Discussion}\label{conclusions}
This  paper has  presented the  results of  a year  of  monitoring the
gravitationally  lensed, ultraluminous  broad  absorption line  quasar
\apm. Photometry relative to several  stars in the field reveals that,
in the  latter months of  these observations, \apm\ has  brightened by
$\sim0.2$ magnitudes.  There are  two potential causes for the source
of this variability; either this  high redshift quasar source is being
microlensed by  stars in  a foreground galaxy,  or the  variability is
intrinsic to  the quasar source.  Both hypotheses  are consistent with
the current data and further photometric monitoring is required before
the exact source of the variability can be determined.

Evidence for gravitational microlensing in \apm\ raises an interesting
possibility. As discussed earlier,  even accounting for the effects of
macrolensing the luminosity of  \apm\ is substantial and intrinsically
it still  ranks amongst  the most luminous  objects known.   But could
\apm\ be a `normal' quasar that is undergoing an extreme gravitational
microlensing magnification  event?  Two arguments suggest  this is not
the case:
\begin{itemize}
\item{\bf  Differential Lensing:}  The  degree to  which  a source  is
magnified is dependent  upon its scale size, with  small sources being
more  strongly magnified~\citep{chang84}.  During the  microlensing of
quasars this results in the enhancement of only the continuum emission
from  the central  accretion disk,  with  the emission  from the  more
extended      broad      line      region     remaining      unchanged
[c.f.~\citet{lewis98a}].   Similarly,   in  \apm,  while   the  quasar
continuum flux  could be enhanced by microlensing,  the sub-mm/IR flux
which  arises  in an  extended  dusty  region  will remain  unchanged.
Subject   to  only  macrolensing   effects,  the   inferred  sub-mm/IR
luminosity of \apm\ is  a prodigious $\sim10^{13}{\rm\lsun}$ (Ibata et
al. 1999; Egami et al. 1999) and it remains intrinsically an extremely
luminous object.
\item{\bf  Statistical Considerations:}  Microlensing is  a stochastic
process whose  probability of  magnification, $\mu$, falls  rapidly at
high values (${p(\mu)\propto \mu^{-3}\  \ \mu \gg 1}$).  Examining the
high  resolution images~\citep{ibata99,egami99},  the two  main images
(A\&B) are of very similar  brightness, within 25\% of each other.  If
we  were  (un)lucky  enough  to   have  caught  \apm\  during  a  high
magnification  event,   high  enough  to  account   for  its  apparent
hyperluminous  properties, is  it possible  that more  than  one image
would be in such a state?  If  we assume that to be in this apparently
ultraluminous state  an image suffers a  microlensing amplification of
greater  than $\sim100$, then  the probability  of finding  the second
image also magnified by this factor is less than $\sim10^{-8}$.  Hence
it  is very unlikely  that both  images are  simultaneously undergoing
substantial microlensing enhancement.
\end{itemize}
We conclude,  therefore, that while  microlensing may account  for the
observed variability, it cannot play a dominant role in distorting our
view. Intrinsically \apm\ must be an extremely luminous system.

The identification of variability in \apm\ can further help reveal the
true nature of  this complex system. High resolution  images with both
NICMOS on the Hubble  Space Telescope~\citep{ibata99} and with the 10m
Keck Telescope~\citep{egami99} clearly show that the two bright images
identified in earlier observations  are accompanied by a fainter image
(designated C) located between  them.  While several arguments suggest
that this  is a third  image of the  high redshift quasar  source, its
location is suspiciously  close to where one would  expect to find the
foreground  lensing  galaxy~\citep{ibata99}.   If  the  high  redshift
source  is  indeed  variable,  a high  spatial  resolution  monitoring
program would  clearly reveal whether C  also varies in  step with the
other images  (subject to a gravitational lens  time delay), revealing
it to truly  be a third image, or the  lensing galaxy.  This procedure
would  also  determine the  degree  of  microlensing  in each  of  the
individual images.

While  current   observations  are  unable  to   fully  constrain  the
gravitational lensing geometry in \apm, the simple image configuration
suggests that the  dominant mass along the line of  sight is a single,
isolated  galaxy (Ibata  et al.  1999;  Egami et  al. 1999).   Further
observations,  such as  deep imaging  of the  field around  \apm, will
refine this picture as it will allow an accurate model for the lensing
potential  to   be  determined.   A  variable  source   and  a  simple
gravitational lensing  geometry suggests that \apm\  may represent one
of  the  few `golden'  gravitational  lenses  from which  cosmological
parameters  can be derived  from time  delay measurements  between the
various images~\citep{williams97,lovell98}.  \apm\ is clearly an ideal
subject for a high resolution monitoring campaign.

\acknowledgments 
We wish to  thank M. Hudson for helpful  discussions and the anonymous
referee for constructive comments.

\newpage

\begin{figure*}
\centerline{
\psfig{figure=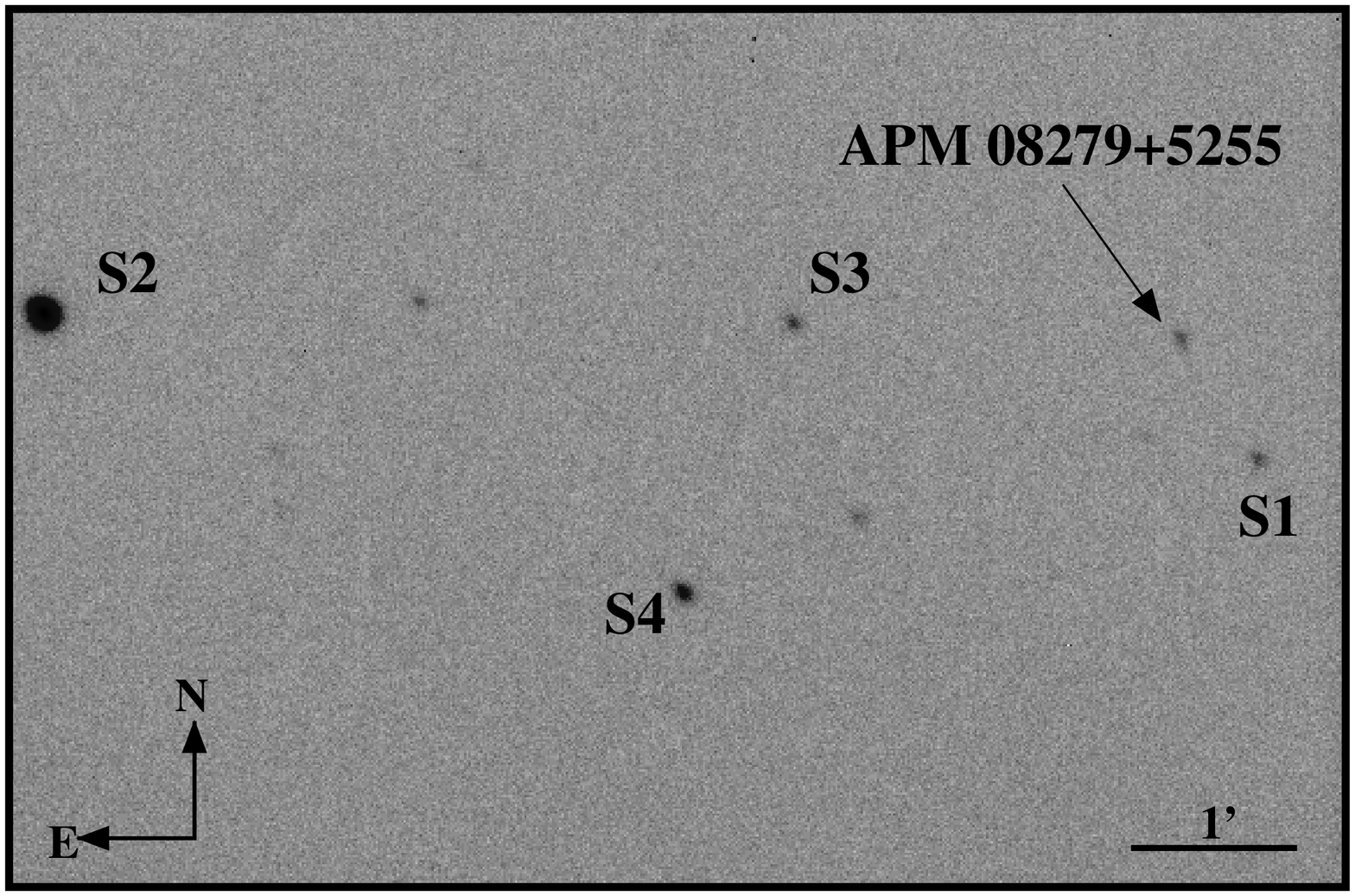,angle=270,width=4in}
}
\caption[]{The observed  field of \apm\  as observed in the  R-band by
the 0.5m telescope of the  Climenhaga Observatory at the University of
Victoria.  As well as the quasar, several comparison stars are marked;
{\bf S1}  is close to \apm\  and is of similar  brightness, while {\bf
S2}  is considerably  brighter than  \apm, allowing  for  more precise
photometry.   Two other stars  which are  employed in  the photometric
analysis ({\bf S3} \& {\bf S4}) are also denoted.  }
\label{figure1}
\end{figure*}

\begin{figure*}
\centerline{
\psfig{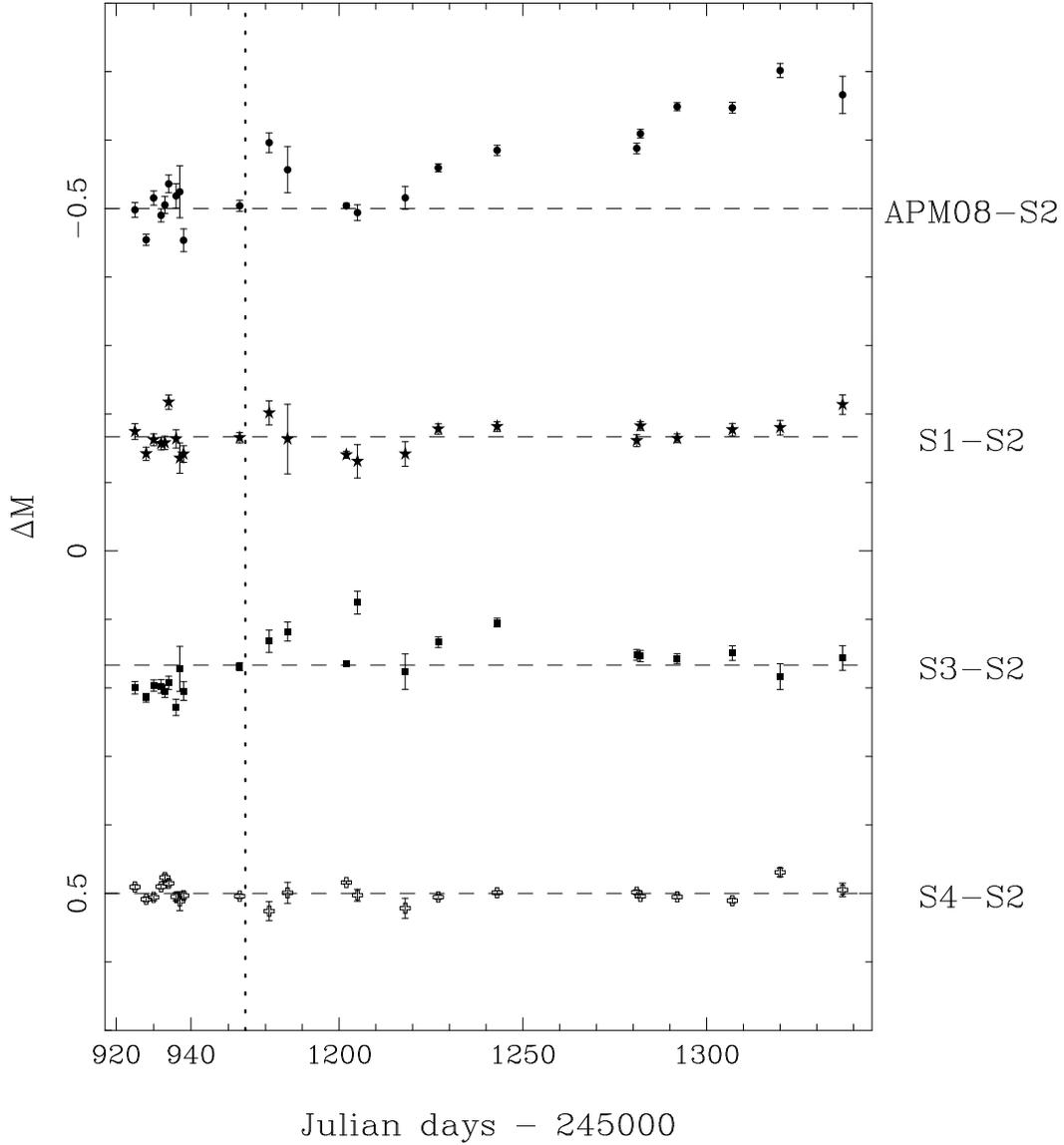}
}
\caption[]{The differential R-Band light curves of the objects denoted
in Figure~1 as compared to the  bright star {\bf S2}. The light curves
have been offset to improve  clarity. For each light curve the offsets
are;  (APM08-{\bf  S2},-4.548),   ({\bf  S1}-{\bf  S2},-4.095),  ({\bf
S3}-{\bf S2},-3.648)  \& ({\bf  S4}-{\bf S2},-2.405).  The  error bars
represent  the uncertainty  in the  mean value  of combining  a single
night  of data,  and the  vertical line  delineates the  two observing
seasons.  }
\label{figure2} 
\end{figure*} 

\newpage

%
%
\begin{deluxetable}{cccccc}
\tablecolumns{6}
\tablecaption{Summary of the differential R-band photometry}
\tablewidth{0pc} 
\tablehead{\colhead{Date}&\colhead{${\rm N_{frame}}$}&\colhead {\bf APM08-S2}&\colhead{\bf S1-S2}&\colhead{\bf S3-S2}&\colhead{\bf S4-S2}}
\startdata
 925& 56&${4.049\pm0.011}$&${3.920\pm0.012}$&${3.847\pm0.009}$&${2.896\pm0.004}$\\
 928& 79&${4.093\pm0.008}$&${3.953\pm0.010}$&${3.862\pm0.006}$&${2.914\pm0.004}$\\
 930& 87&${4.032\pm0.010}$&${3.932\pm0.009}$&${3.844\pm0.008}$&${2.911\pm0.005}$\\
 932& 48&${4.057\pm0.009}$&${3.938\pm0.009}$&${3.845\pm0.010}$&${2.896\pm0.004}$\\
 933& 33&${4.042\pm0.013}$&${3.937\pm0.010}$&${3.852\pm0.009}$&${2.882\pm0.005}$\\
 934& 74&${4.012\pm0.013}$&${3.877\pm0.010}$&${3.840\pm0.010}$&${2.891\pm0.005}$\\
 936& 68&${4.029\pm0.018}$&${3.931\pm0.013}$&${3.876\pm0.012}$&${2.910\pm0.006}$\\
 937&  9&${4.023\pm0.038}$&${3.959\pm0.022}$&${3.820\pm0.033}$&${2.917\pm0.014}$\\
 938& 20&${4.094\pm0.017}$&${3.953\pm0.012}$&${3.852\pm0.014}$&${2.909\pm0.005}$\\
 953& 66&${4.043\pm0.008}$&${3.929\pm0.008}$&${3.816\pm0.006}$&${2.909\pm0.003}$\\
1181& 18&${3.951\pm0.014}$&${3.893\pm0.018}$&${3.779\pm0.016}$&${2.931\pm0.014}$\\
1186&  2&${3.991\pm0.034}$&${3.931\pm0.051}$&${3.765\pm0.014}$&${2.905\pm0.015}$\\
1202&101&${4.043\pm0.004}$&${3.954\pm0.005}$&${3.812\pm0.003}$&${2.889\pm0.002}$\\
1205& 13&${4.054\pm0.011}$&${3.964\pm0.024}$&${3.723\pm0.017}$&${2.908\pm0.009}$\\
1218& 10&${4.032\pm0.017}$&${3.953\pm0.018}$&${3.824\pm0.026}$&${2.927\pm0.015}$\\
1227& 47&${3.988\pm0.006}$&${3.916\pm0.008}$&${3.781\pm0.008}$&${2.911\pm0.003}$\\
1243& 48&${3.963\pm0.007}$&${3.913\pm0.007}$&${3.752\pm0.006}$&${2.904\pm0.003}$\\
1281& 27&${3.960\pm0.008}$&${3.933\pm0.009}$&${3.799\pm0.008}$&${2.904\pm0.004}$\\
1282& 93&${3.938\pm0.006}$&${3.912\pm0.006}$&${3.801\pm0.008}$&${2.910\pm0.003}$\\
1292& 93&${3.899\pm0.006}$&${3.930\pm0.007}$&${3.805\pm0.007}$&${2.910\pm0.003}$\\
1307& 70&${3.900\pm0.008}$&${3.917\pm0.009}$&${3.797\pm0.011}$&${2.916\pm0.004}$\\
1320& 22&${3.846\pm0.010}$&${3.914\pm0.011}$&${3.831\pm0.019}$&${2.875\pm0.006}$\\
1337& 29&${3.881\pm0.027}$&${3.881\pm0.014}$&${3.804\pm0.018}$&${2.900\pm0.010}$\\
\enddata
\tablecomments{
Summary of the R-band differential  magnitudes obtained on each of the
objects denoted in Figure~1, with respect to the bright star {\bf S2}.
Here, the  date of the observation  is given in Julian  Days - 245000,
while ${\rm N_{frame}}$  is the number of images  obtained that night.
The errors represent the uncertainty in the mean of the combination of
these frames.}
\end{deluxetable}

\end{document}